\documentclass[a4paper,11pt]{article}
\usepackage{pos}
\usepackage{lineno}

\title{Theoretical derivation of diffusion-tensor coefficients for the transport of charged particles in magnetic fields}

\author[]{Olivier Deligny}

\affiliation[]{Laboratoire de Physique des 2 Infinis Ir\`ene Joliot-Curie (IJCLab)\\
CNRS/IN2P3, Universit\'{e} Paris-Saclay, Orsay, France}

\emailAdd{deligny@ijclab.in2p3.fr}

\abstract{The transport of charged particles in various astrophysical environments permeated by magnetic fields is described in terms of a diffusion process, which relies on diffusion-tensor parameters generally inferred from Monte-Carlo simulations. Based on a red-noise approximation to model the two-point correlation function of the magnetic field experienced by charged particles between two successive times, the diffusion-tensor coefficients were previously derived in the case of pure turbulence. In this contribution to ECRS2022, the derivation is extended to the case of a mean field on top of the turbulence. The results are  applicable to a variety of astrophysical environments in regimes where the Larmor radius of the particles is resonant with the power spectrum of the turbulence wavelength (gyro-resonant regime), or where the Larmor radius is greater than the largest turbulence wavelength (high-rigidity regime).}

\FullConference{%
  *** 27th European Cosmic Ray Symposium - ECRS ***\\
  *** 25-29 July 2022 ***\\
  *** Nijmegen, the Netherlands ***
}

\newcommand{\dif}{\mathrm{d}}

\begin{document}
\maketitle


In many astrophysical environments, the propagation and acceleration of high-energy charged particles (cosmic rays) are governed by the scattering off magnetic fields. The transport of the particles is then modelled as an anisotropic diffusion process. Under very broad conditions, the coefficients of the diffusion tensor can be related to the velocity correlation function of cosmic rays, $\langle v_{0i}v_{j}(t)\rangle$, through a time integration~\citep{Kubo:1957mj}, 
\begin{equation}
\label{eqn:Dij-Kubo}
    D_{ij}(t)=\int_0^{t}\dif t'~\langle v_{0i}v_{j}(t')\rangle,
\end{equation}
in the limit that $t\rightarrow\infty$. Here, $v_{0i}\equiv v_i(t=0)$\footnote{Since cosmic rays are high-energy relativistic particles, the norm of the velocity is identified to $c$ for convenience. } and $\langle\cdot\rangle$ stands for the average quantities, taken over several space and time correlation scales of the turbulent field. Many estimates of these coefficients have been made from numerical simulations exploring wide ranges of particle rigidities and turbulence levels~\citep[e.g.][]{1999ApJ...520..204G,PhysRevD.65.023002,Candia:2004yz,2010ApJ...711..997H,Fraschetti:2012cm,Fatuzzo:2014hua,Snodin:2015fza,Reichherzer:2019dmb,Reichherzer:2021yyd}. In this contribution, we present the main steps of a theoretical derivation of the velocity correlation functions consistent with simulation results in a range of rigidities gyro-resonant with the power spectrum of turbulence. Hence, this extends  results obtained in~\cite{Plotnikov:2011me} in the high rigidity regime, and those in~\cite{Deligny:2021wyi} in the gyro-resonant regime limited to pure turbulence. Without loss of generalities, the study is limited to the example of an isotropic 3D turbulence following a Kolmogorov power spectrum without helicity denoted as $\delta\mathbf{B}$, while the mean field is denoted as $\mathbf{B}_0=B_0\mathbf{u}_z$. The turbulence level is defined as $\eta=\delta B^2/(\delta B^2+B_0^2)$.

Due to the stochastic nature of the turbulence, the velocity of the particles is a stochastic variable as well. We are thus interested in determining the moments of $v_i(t)$ from the Lorentz-Newton equation of motion for the particles,
\begin{equation}
    \label{eqn:LorentzNewton}
    \dot{v}_i(t)=\delta\Omega(t) ~\epsilon_{ijk}v_j(t)\delta b_k(t)+\Omega_0 ~\epsilon_{ijk}v_j(t) b_{0k}(t).
\end{equation}
Here, $\delta\Omega(t)=c^2Z|e|\delta B(t)/E$ is the gyrofrequency related to the turbulence, $\Omega_0$ that related to the mean field, $Z|e|$ the electric charge, $E$ the energy of the particle, and $\delta b_k(t)\equiv\delta b_k(\mathbf{x}(t))$ the $k$-th component of the turbulence (expressed in units of $\delta B$) at the spatial coordinate $\mathbf{x}(t)$ of the particle at time $t$. A formal solution for $\langle v_{i}(t)\rangle$ can be obtained by expressing the solution of Eqn.~\ref{eqn:LorentzNewton} as an infinite number of Dyson series, each combining terms in powers of $\delta \mathbf{b}$ coupled to terms in powers of $\mathbf{B}_0$. 

Dealing with such an infinite number of Dyson series is however hardly manageable. To circumvent this, we use the auxiliary variable introduced in~\cite{Plotnikov:2011me}, $w_i(t)=R_{ij}(t)v_j(t)$, with $\mathbf{R}(t)$ the rotation matrix of angle $\Omega_0 t$ around $\mathbf{u}_z$. The equation of motion for $\mathbf{w}$ is then
\begin{equation}
    \label{eqn:LorentzNewton-w}
    \dot{w}_i(t)=\delta\Omega(t)R^{-1}_{ij}\epsilon_{jkl} \delta b_l(t)R_{km}w_m(t),
\end{equation}
the formal solution of which can be expressed as a \textit{single} Dyson series: 
\begin{eqnarray}
    \label{eqn:dyson}
    \langle w_{i_0}(t)\rangle&=&w_{0i_0}+\sum_{p=1}^\infty \delta\Omega^p ~ \epsilon_{k_1m_1n_1}\epsilon_{k_2m_2n_2}\dots\epsilon_{k_pm_pn_p} w_{0i_p} \int_0^t \hspace{-0.25cm}\dif t_1\int_0^{t_1}\hspace{-0.35cm}\dif t_2\dots\int_0^{t_{p-1}}\hspace{-0.60cm}\dif t_p \nonumber \\ 
    &\times&  R^{-1}_{i_0k_1}(t_1)R^{-1}_{i_1k_2}(t_2)\cdots R^{-1}_{i_{p-1}k_p}(t_p)R_{m_1i_1}(t_1)\cdots R_{m_pi_p}(t_p)\langle\delta b_{n_1}(t_1)\dots\delta b_{n_p}(t_p)\rangle.
\end{eqnarray} 
In the following, we derive the velocity correlation function parallel to the mean field, $\langle v_{0z}v_z(t)\rangle$, based on this equation. The perpendicular (and anti-symmetric) functions can be obtained in a similar way and will be detailed in a forthcoming publication~\cite{Deligny:2023}.

In the Gaussian approximation, the Wick theorem allows for expressing the expectation value $\langle\delta b_{n_1}(t_1)\dots\delta b_{n_p}(t_p)\rangle$ in terms of all possible permutations of products of contractions of pairs of $\langle \delta b_{n_i}(t_{i})\delta b_{n_j}(t_{j})\rangle$, which can be, in the case of 3D isotropic turbulence, written as
\begin{equation}
\label{eqn:dbdb}
\langle \delta b_{n_1}(t_{i})\delta b_{n_2}(t_{j})\rangle=\frac{\delta_{n_1n_2}}{3}\varphi(t_{i}-t_{j}).
\end{equation}
The correlation function $\varphi(t)$, which describes the correlation of the turbulence experienced by a particle along its path at two different times, is modeled in this study as a red-noise process with parameter $\tau$, 
\begin{equation}
\label{eqn:phi}
\varphi(t)=\exp{(-t/\tau)}.
\end{equation}
The expression of $\tau$, similarly to that found in~\cite{PhysRevD.65.023002}, depends on the regime of rigidity considered. With $\rho$ the Larmor radius of the particle expressed in units of the largest eddy scale $L_{\mathrm{max}}$ of the turbulence, $\tau\simeq L_\mathrm{c}/c$ for $\rho\gtrsim \pi L_{\mathrm{c}}/L_{\mathrm{max}}$, where $L_{\mathrm{c}}$ is the coherence scale of the turbulence. This is because in this rigidity regime, particles can travel over a distance $L_\mathrm{c}$ undergoing small deflections only. On the other hand, for $\rho \lesssim \pi L_{\mathrm{c}}/L_{\mathrm{max}}$, the following heuristic estimate that makes use of the kinetic energy spectrum of the turbulence $\mathcal{E}(k)$ is observed to reproduce simulations:
\begin{equation}
    \label{eqn:tau}
    \tau\simeq\frac{2}{c}\frac{\int_{k_\star}^{k_{\mathrm{max}}}\dif k k^{-1}\mathcal{E}(k)}{\int_{k_\star}^{k_{\mathrm{max}}}\dif k \mathcal{E}(k)}.
\end{equation}
In this regime of rigidity, $\tau$ inherits a $\rho$ dependency from that of the lower boundary in wave number $k_\star(\rho)=\rho_\star k_{\mathrm{min}}/\rho$ with $\rho_\star=2L_{\mathrm{c}}/(\pi L_{\mathrm{max}})$. The truncation in the wavenumber integration range selects modes for which particles do not experience spiral motions around the corresponding large-scale magnetic field lines over several Larmor times, modes that hence prevent decorrelations from occurring on relevant time scales.

To carry out a summation of the infinite Dyson series (Eqn.~\ref{eqn:dyson}), we resort to a two-step iteration procedure. First, the same partial summation scheme as in~\cite{Deligny:2021wyi} is used. The classes of ``diagrams'' retained in the scheme are of two kinds: unconnected and nested ones~\cite{1966AnAp...29..645F}. This corresponds formally to the Kraichnan propagator~\cite{Kraichnan:1961:DNS}. In fact, it has been shown that the summation of the first two terms (beyond the free propagator) is sufficient to give us a physical solution in the case of pure turbulence~\cite{Deligny:2021wyi}. We follow here the same strategy so that, making use of the properties of the Levi-Civita symbols contracted over one index,  Eqn.~\ref{eqn:dyson} is substituted by
\begin{eqnarray}
    \label{eqn:dysonK0}
    \langle w^\mathrm{K}_{z}(t)\rangle&=&w_{0z}+\frac{\delta\Omega^2}{3} \int_0^t \hspace{-0.25cm}\dif t_1\int_0^{t_1}\hspace{-0.35cm}\dif t_2~R^{-1}_{zk}(t_1)R_{mi_1}(t_1)R^{-1}_{i_1k}(t_2)R_{mi_2}(t_2)\varphi(t_1-t_2) \langle w^\mathrm{K}_{z}(t-t_1)\rangle  \nonumber \\ &-&\frac{\delta\Omega^2}{3} \int_0^t \hspace{-0.25cm}\dif t_1\int_0^{t_1}\hspace{-0.35cm}\dif t_2~R^{-1}_{zk_1}(t_1)R_{k_2i_1}(t_1)R^{-1}_{i_1k_2}(t_2)R_{k_1i_2}(t_2)\varphi(t_1-t_2) \langle w^\mathrm{K}_{z}(t-t_1)\rangle  \nonumber \\
    &+& \left(\frac{\delta\Omega^2}{3}\right)^2 \int_0^t \hspace{-0.25cm}\dif t_1\cdots\int_0^{t_3}\hspace{-0.35cm}\dif t_4~\varphi(t_1-t_4)\varphi(t_2-t_3) \langle w^\mathrm{K}_{z}(t-t_1)\rangle \times \nonumber \\
    && (R^{-1}_{zk_1}(t_1)R_{m_1i_1}(t_1)R^{-1}_{i_1k_2}(t_2)R_{m_2i_2}(t_2)R^{-1}_{i_2k_2}(t_3)R_{m_2i_3}(t_3)R^{-1}_{i_3k_1}(t_4)R_{m_1z}(t_4)  \nonumber \\
    &-&R^{-1}_{zk_1}(t_1)R_{m_1i_1}(t_1)R^{-1}_{i_1k_2}(t_2)R_{m_2i_2}(t_2)R^{-1}_{i_2m_2}(t_3)R_{k_2i_3}(t_3)R^{-1}_{i_3k_1}(t_4)R_{m_1z}(t_4) \nonumber \\
    &-&R^{-1}_{zk_1}(t_1)R_{m_1i_1}(t_1)R^{-1}_{i_1k_2}(t_2)R_{m_2i_2}(t_2)R^{-1}_{i_2k_2}(t_3)R_{m_2i_3}(t_3)R^{-1}_{i_3m_1}(t_4)R_{k_1z}(t_4) \nonumber \\
    &+&R^{-1}_{zk_1}(t_1)R_{m_1i_1}(t_1)R^{-1}_{i_1k_2}(t_2)R_{m_2i_2}(t_2)R^{-1}_{i_2m_2}(t_3)R_{k_2i_3}(t_3)R^{-1}_{i_3m_1}(t_4)R_{k_1z}(t_4)),
\end{eqnarray} 
where the superscript K stands for ``Kraichnan''. Next, some properties of the rotation matrices, in particular $R^{-1}_{zk}=R_{zk}=\delta_{zk}$,  $R_{ij}(t_1)R^{-1}_{jk}(t_2)=R_{ik}(t_1-t_2)$, and $R_{ij}(t_1-t_2)=R_{ji}(t_2-t_1)$), allow us to get an explicit form of the equation:
\begin{eqnarray}
    \label{eqn:dysonK}
    \langle w^\mathrm{K}_{z}(t)\rangle&=&w_{0z}- \frac{\delta\Omega^2}{3} \int_0^t \hspace{-0.25cm}\dif t_1\int_0^{t_1}\hspace{-0.35cm}\dif t_2~\varphi(t_1-t_2) \cos{(\Omega_0(t_1-t_2))}\langle w^\mathrm{K}_{z}(t-t_1)\rangle  \nonumber \\ 
    &+& 4\left(\frac{\delta\Omega^2}{3}\right)^2 \int_0^t \hspace{-0.25cm}\dif t_1\cdots\int_0^{t_3}\hspace{-0.35cm}\dif t_4~\varphi(t_1-t_4)\varphi(t_2-t_3)\cos{(\Omega_0(t_2-t_3))}\cos{(\Omega_0(t_1-t_2+t_3-t_4))}\langle w^\mathrm{K}_{z}(t-t_1)\rangle \nonumber \\
    &-& 2\left(\frac{\delta\Omega^2}{3}\right)^2 \int_0^t \hspace{-0.25cm}\dif t_1\cdots\int_0^{t_3}\hspace{-0.35cm}\dif t_4~\varphi(t_1-t_4)\varphi(t_2-t_3)\cos{(\Omega_0(t_1-2t_2+2t_3-t_4))}\langle w^\mathrm{K}_{z}(t-t_1)\rangle \nonumber \\ 
    &+& 2\left(\frac{\delta\Omega^2}{3}\right)^2 \int_0^t \hspace{-0.25cm}\dif t_1\cdots\int_0^{t_3}\hspace{-0.35cm}\dif t_4~\varphi(t_1-t_4)\varphi(t_2-t_3)\cos{(\Omega_0(t_1-t_2+t_3-t_4))}\langle w^\mathrm{K}_{z}(t-t_1)\rangle.
\end{eqnarray} 
To solve this non-linear equation, we proceed with a Laplace transform. The change of variables $t=x+x_1+\cdots +x_p$, $t_1=x_1+\cdots +x_p$, $t_2=x_2+\cdots +x_p$,$\cdots$, $t_p=x_p$ allows for sending all integration boundaries between 0 and $+\infty$ for the $x_i$ variables. After some algebra, the equation for the Laplace transform $\mathcal{L}\left(w(t)\right) \equiv \hat{W}(p)$ reads as
\begin{eqnarray}
    \label{eqn:dysonK-Laplace}
    \hat{W}^\mathrm{K}(p)&=&\frac{1}{p}- 2\frac{\delta\Omega^2}{3}\frac{\hat{W}^\mathrm{K}(p)}{p}\frac{p+\tau^{-1}}{(p+\tau^{-1})^2+\Omega_0^2} \nonumber \\
    &+& 2\left(\frac{\delta\Omega^2}{3}\right)^2\frac{\hat{W}^\mathrm{K}(p)}{p}\left(\frac{p+2\tau^{-1}}{(p+2\tau^{-1})^2+\Omega_0^2}\right)\left(\frac{(p+\tau^{-1})^2-\Omega_0^2}{\left((p+\tau^{-1})^2+\Omega_0^2\right)^2}\right)   \nonumber \\
    &+& 2\left(\frac{\delta\Omega^2}{3}\right)^2\frac{\hat{W}^\mathrm{K}(p)}{p}\left(\frac{1}{p+2\tau^{-1}}\right)\left(\frac{(p+\tau^{-1})^2-\Omega_0^2}{\left((p+\tau^{-1})^2+\Omega_0^2\right)^2}\right)   \nonumber \\
    &-& 4\left(\frac{\delta\Omega^2}{3}\right)^2\frac{\hat{W}^\mathrm{K}(p)}{p}\left(\frac{\Omega_0^2(p+\tau^{-1})}{\left((p+2\tau^{-1})^2+\Omega_0^2\right)\left((p+\tau^{-1})^2+\Omega_0^2\right)^2}\right).
\end{eqnarray} 
The solution for $\langle w^\mathrm{K}_z(t)\rangle$ is then obtained by making use of the numerical Stehfest scheme of the inverse Laplace transform. 

Once $\langle w^\mathrm{K}_z(t)\rangle$ is determined, the second step to get an improved estimation of the propagator consists in including ``crossed'' diagrams (that formally account for any mix of crossed and nested diagrams). To do so, an iterative procedure is designed to carry out this summation. In the Laplace space, the $N$th iterated $\langle w^\mathrm{N}_z(t)\rangle$ reads as~\cite{Deligny:2023}
\begin{eqnarray}
    \label{eqn:dysonK-Laplace-N}
    \hat{W}^N(p)&=&\frac{1}{p}- 2\frac{\delta\Omega^2}{3}\frac{\hat{W}^N(p)}{p}\mathcal{L}\left(\exp{(-t/\tau)w^{N-1}(t)\cos{\Omega_0t}}\right) \nonumber \\
    &+& 2\left(\frac{\delta\Omega^2}{3}\right)^2\frac{\hat{W}^N(p)}{p}\mathcal{L}\left(\exp{(-2t/\tau)w^{N-1}(t)}\right)\mathcal{L}^2\left(\exp{(-t/\tau)w^{N-1}(t)\cos{\Omega_0t}}\right),
\end{eqnarray} 
with the initial iteration $w^0(t)=w^\mathrm{K}(t)$. In practice, convergence is achieved after a few iterations ($N \simeq 5$).

\begin{figure}[t]
\centering
\includegraphics[width=\columnwidth]{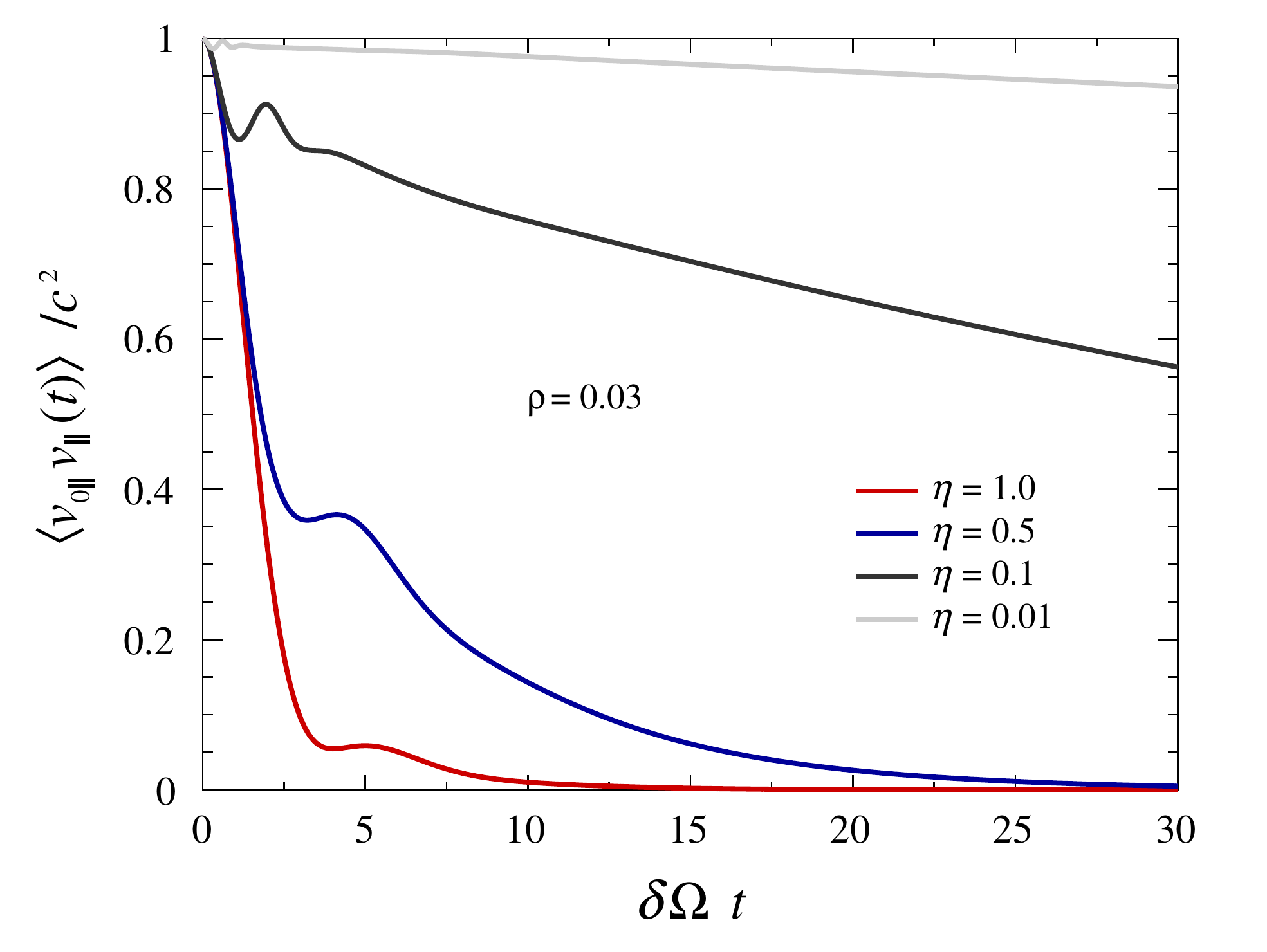}
\caption{Time dependence of the auto-correlation of the
particle velocities parallel to the mean field, for different values of turbulence level. A reduced rigidity $\rho=0.03$ is chosen to illustrate the gyro-resonant regime.}
\label{fig:Rpar}
\end{figure}

The resulting picture is illustrated in Fig.~\ref{fig:Rpar}, where the time dependence of the auto-correlation of the
particle velocities parallel to the mean field is shown for different values of turbulence level. The reduced rigidity is chosen to be $\rho=0.03$ so as to explore the gyro-resonant regime. The timescale of the correlation is observed to be minimal in the case of pure turbulence ($\eta=1$) and to tend to infinity in the case of low turbulence. The various structures beyond the exponential falloff that have been uncovered in most of the Monte-Carlo simulations mentioned in the introduction are reproduced by the calculation presented in this contribution. 


\bibliographystyle{JHEP}
\bibliography{bibliography}


\end{document}